\documentclass[aps, prd, twocolumn, lengthcheck, superscriptaddress, showpacs, letterpaper, nofootinbib]{revtex4-1}

\usepackage{graphicx}
\usepackage{color}

\def\la{\langle}\def\ra{\rangle}
\def\be{\begin{eqnarray}}\def\ee{\end{eqnarray}}
\def\lsim{\mathrel{\rlap{\lower3pt\hbox{\hskip1pt$\sim$}}
     \raise1pt\hbox{$<$}}} 
\def\gsim{\mathrel{\rlap{\lower3pt\hbox{\hskip1pt$\sim$}}
     \raise1pt\hbox{$>$}}} 
\def\le{ \begin{array}{ll}}\def\re{\end{array}}

\def\lear{ \left( \begin{array}{cc}}\def\rear{\end{array} \right)}

\def\le{ \left( \begin{array}{cc}}\def\re{\end{array} \right)}

\def\bi{\bibitem}
\def\del{\partial}

\def\la{\langle}\def\ra{\rangle}
\def\be{\begin{eqnarray}}\def\ee{\end{eqnarray}}
\def\lsim{\mathrel{\rlap{\lower3pt\hbox{\hskip1pt$\sim$}}
     \raise1pt\hbox{$<$}}} 
\def\gsim{\mathrel{\rlap{\lower3pt\hbox{\hskip1pt$\sim$}}
     \raise1pt\hbox{$>$}}} 
\def\etaprime{\eta^\prime} 
\def\FSB{{\cal B}^S}

\begin{document}

\title{Probing Fractional Quantum Hall Sheets in Dense Baryonic Matter}

\author{Mannque Rho}

\affiliation{
Universit\'e Paris-Saclay, Institut de Physique Th\'eorique,  CNRS, 91191 Gif-sur-Yvette c\'edex, France
}

\date{\today}

\begin{abstract}
Unlike the octet baryons for $N_f=3$,  there is no skyrmion coming from the $\eta^\prime$ meson. It is instead described as a fractional quantum Hall droplet, a pancake or a pita involving a singular $\eta^\prime$ ring in which Chern-Simons fields live. By incorporating hidden local symmetry and hidden scale symmetry in nuclear dynamics, I describe how to access baryon-charged quantum Hall droplets in dense nuclear matter in terms of the nuclear scale-chiral effective field theory approach ``G$n$EFT" with the $U_A(1)$ anomaly  taken into account. I discuss how  the single-flavor baryon that I will call ${\cal B}^s$ could be exposed in superdense baryonic matter,  figuring, perhaps, in  ``quark stars" associated with  the baryon-quark continuity involving  CFL or    phase transitions.
 \end{abstract}


\maketitle



\section{Introduction}
Treating baryons as skyrmions~\cite{skyrme} in the large $N_c$ limit is considered to be a {\it first-principles} approach to  nuclear physics. In contrast with the power and spectacular success of the skyrmion backed by experiments in condensed matter physics in dimensions $D=d+1$ less than 4\footnote{Partial developments up to 2017 of skyrmions in condensed matter, nuclear, particle physics and in string theory are found in  \cite{rho-zahed}. Some impressive breakthrough has been made since then.},  it has been found immensely difficult to work out in practice the skyrmion model to confront nuclear dynamics where it was originally envisioned by Skyrme, whereas the nuclear chiral effective field theory (N$\chi$EFT for short)~\cite{chieft}, another approach heralded as ``first-principles" approach,  has proven to fare fairly well at densities up  to nuclear matter one,  $n_0\sim 0.16$ fm$^{-3}$.   Apart from the fact that the {\it effective} Lagrangian from which the skyrmion structure should be deduced is not even adequately known, the mathematics involved in its treatment has been a major stumbling block to exploiting the topological structure of the theory  in nuclear dynamics. 

Recently however the situation has changed in a significant way.  

There are two notable developments for this, one in accessing the structure of light nuclei with mass number $A\lsim 32$ and the other approaching high densities considered relevant to massive compact stars. 

The former has been achieved, most prominently,  by Manton and his co-workers, as recently reccounted in an elegant monograph volume~\cite{manton-book}. Nuclei ranging from $B=2$ to $B=32$ are treated as topological solitons with the $B=A$ identified as the topological number. The approach consists of resolving the quantal equations of motion of the skyrmion configurations,  resorting to powerful ans\"atze.  In \cite{manton-book}, the ``rational map" is successfully exploited.   The strategy adopted in this approach is quite distinct from the standard nuclear theory approaches that resort to nucleon-nucleon interaction potentials,  average potentials resorting to such notions as shell structure, Fermi sea, etc., that  in condensed matter language could be classified as  quantum critical phenomena. Some of the results obtained with the original -- and simplest -- Skyrme model with pions only are in surprisingly good agreement with Nature.  One could expect, with additional degrees of freedom such as the tower of hidden symmetry bosons as indicated in holographic dual QCD models taken into consideration,  to be able to access mass numbers $B > 32$ and dense matter of  compact stars.

The second issue, i.e.,  going to heavier nuclei and, specially, dense matter, which in the direct skyrmion approach, even with the simplest Skyrme model with no additional fields than the pions, has been lacking. The mathematics is too daunting in many-baryon systems. However it has been feasible to exploit certain topological power of the skyrmion structure to build a nuclear effective field theory, dubbed G$n$EFT, that enables one to go from the normal density $\sim n_0$  to $\sim (4-7) n_0$, relevant to neutron stars, at which the N$\chi$EFT is to break down.  The basic idea, amply reviewed~\cite{MR-reviews}, is to inject the topological prowess into an effective Lagrangian that captures hidden symmetries associated with ``heavy degrees of freedom (HdF)." It has been found to explain -- so far with no tensions with  -- the presently available data in compact-star observations~\cite{lattimer}.

What I will describe here addresses {\it what is not} in the G$n$EFT up to the densities probed in compact stars, namely, the possible topological object associated with $\etaprime$.   In the large $N_c$ limit, $\etaprime$ will be as light as the pions.  So as mentioned below, $\etaprime$ is expected to play a role in nuclear physics. But unlike the octet baryons that come as skyrmions from the octet mesons, $\etaprime$ does not give a skyrmion description. One of the recent developments is that there is indeed  a topological object involving a singular $\etaprime$ configuration carrying  a baryon charge,  called  ``quantum Hall droplet" involving Chern-Simons topological fields  that is a topological object coming from $\etaprime$~\cite{komargodski}. I will call  it  $\FSB$. In contrast to $\etaprime$ which is a flavor--singlet meson, it is not a favor-singlet baryon\footnote{The flavor singlet $uds$-quark baryon configuration is one of $\Lambda_s$'s. $\FSB$  is what appears to be a  ``fictitious"  single-flavor ($N_f=1$) baryon with $J^\pi=3/2^+$, not listed in the Particle Data Booklet. Unlike $\Delta (3/2,3/2)$ which has the same quantum numbers, its mass has a term $\propto O(N_c^0)$. }. 

Now given that it is not a skyrmion, how does it manifest in nuclear matter which in skyrmion picture is a two-flavor -- up-quark and down-quark -- system in QCD ?  This is a relevant physical issue.  This is because the light-quark flavor singlet meson $\etaprime$ has been extensively studied, starting with  chiral symmetry restoration at high temperature~\cite{prodigalGB}. Given that the $\etaprime$ mass, $\sim 960$ MeV in matter-free space, could drop substantially in nuclear medium, say, $\sim 100$  MeV, at nuclear matter density, there is a growing effort to study at various laboratories what happens to $\etaprime$s in nuclear matter, for example, the $\etaprime$ energy spectra for the $\etaprime$ mesonic nuclei in $^{12}$C, $^{16}$O etc.~\cite{jidoetal}. Now if the $\etaprime$ begins to be bound in nuclear medium, the question then would be what about the $\FSB$s? To start with, how does one go about addressing this question?

It is the objective of this article to explore the possibility of probing $\FSB$s as quantum Hall droplets in dense baryonic matter. As is generally believed. the natural setting to treat dense baryonic matter in QCD is the skyrmion matter~\cite{park-vento}. Therefore one would like to start with skyrmion matter and ``drive" the system by distorting the skyrmion configurations to that of a $\FSB$. 

One way to drive a skyrmion matter to  $\FSB$ matter was discussed in \cite{karasik1}.  Suppose one starts with $N_f=2$ (that could  be generalized to $N_f >  2$) with $m_u=0$ and $m_d=0$ and  the chiral field with $\etaprime$ included as
$U=e^{i\etaprime} e^{i\pi/f_\pi}$. The nuclei and infinite matter described in terms of skyrmions, it seems highly reasonable, would be unaffected by the presence of $\etaprime$. Now do what one might call  ``toy-model" manipulation to increase  $m_d$ toward $\infty$ while keeping $m_u=0$.  It has been shown~\cite{karasik1} that as $m_d$ is increased smoothly to $\infty$, the hedgehog configuration of the skyrmion gets distorted in such a way that the skyrmion transforms to a configuration in which $\etaprime$ winds around a singular ring with $\etaprime$ jumping from $\etaprime=-\pi$ to $\etaprime=\pi$ as one crosses the sheets. This represents the $\FSB$.  What jumps across is identified with ``heavy degrees of freedom (HdF)," which are gluons in QCD and hidden gauge bosons by (Seiberg-)duality~\cite{karasik1}.

Obviously this toy-model manipulation of the mass $m_d$ is physically unrealizable. What I propose as a physically accessible means is to drive the usual skyrmion system, i.e., nuclear matter,  by increasing density to some high value. This can be done experimentally in current or upcoming laboratories and possibly in the future observations of  massive compact stars. My strategy is to take the presently workable nuclear EFT,  G$n$EFT, which has been found to satisfactorily explain what's observed in compact-star physics to density $\sim 7 n_0$, and extrapolate it to higher densities taking into account  what was assumed to hold beyond the density regime applicable to compact stars, for instance,  the ``vector manifestation (VM)" of HLS and  the approach to the ``dilaton-limit fixed point (DLFP)" of hidden scale symmetry, both of which have played a key role in compact-star physics so far worked out in G$n$EFT~\cite{MR-reviews}.

\section{Quantum Hall Sheets and Hidden Local Symmetry}
\subsection{$\etaprime$ cusp singularity and heavy degrees of freedom (HdF)}
Because of the established $U_A(1)$ anomaly,  the low-energy effective theory of $\etaprime$ is gapped even in the chiral limit. The mass goes as $1/N_c$, so in the large $N_c$ limit, it has chiral symmetry which breaks down spontaneously, giving rise to a Nambu-Goldstone (NG) boson $\etaprime$.  As is well-known the effective Lagrangian for $\etaprime$
\be
{\cal L}_\eta^\prime=\frac{f_\pi^2}{2}(\del_\mu \etaprime)^2 - V_{\etaprime}
\ee
with the potential
\be
V_{\etaprime} \approx \frac 12 f_\pi^2 m_{\etaprime}^2  min _{k\in Z} (\etaprime - 2\pi k)^2 +\cdots\label{cusp}
\ee
has the cusp singularity whenever $\etaprime=\pi\ {\rm mod}\ 2\pi$ representing a phase transition between two branches. The cusp is generated as ``heavy fields," i.e., gluons in QCD, jump from one vacuum to another vacuum~\cite{vecchia-veneziano,karasik1}. For small fluctuations around the vacuum $\etaprime=0$, it gives a mass term but when certain global effects such as nontrivial windings of $\etaprime$ are present, then the cusp plays an intricate role.  How this leads to the fractional quantum Hall structure on the $\etaprime$ sheet as a configuration carrying a baryon charge that interpolates between $\etaprime=0$ and $\etaprime=2\pi$ captured by a $U(1)_{N_c}$ Chern-Simons theory is explained in detail in \cite{komargodski}.  There is also an indication for such a $\FSB$ in holographic dual QCD~\cite{dualqcd}.

For those like the present author who are not familiar with the language, it might be easier to see what the difference is between the skymion and the quantum Hall droplet in terms of the Cheshire-Cat phenomenon (CCP)~\cite{CCP}. In terms of CCP, the $N_f=2$ skyrmion corresponds to the $N_c$ quarks falling into an ``infinite hotel" whereas the $N_f=1$ quantum Hall droplet corresponds to the quarks doing the Callan-Harvey anomaly outflow~\cite{callan-harvey}  to a sheet carrying Chern-Simons topological field theory. 

There are two recent developments that prompted me to write this article.  The first, principal, one is the proposal by Karasik~\cite{karasik1,karasik2} that the cusp (\ref{cusp}) can be eliminated by the vector mesons in hidden local symmetry (HLS) in the effective theory~\cite{HLS,HY:PR} and the second one is how baryonic system could be driven by high density exploiting  the  ``genuine dilaton (GD)" notion in hidden scale symmetry of Crewther et all.~\cite{CT,GD}. This could offer a means to expose the $\FSB$ if it is present in the baryonic matter.  Precisely these two (hidden) symmetries turned out to play the principal role in the G$n$EFT approach in accessing massive compact stars~\cite{MR-reviews}.

Briefly stated following \cite{karasik1,karasik2}, how the HLS vector mesons figure in the problem goes as follows. As is well-known, the effective theory for the gluonic topological density $Q=\frac{1}{8\pi^2} {\rm tr}\ G\wedge G$ (where $G$ is the gluon field tensor) is given by
\be
{\cal L}_Q= \frac{1}{2f_\pi^2 m^2_{{\eta^\prime}} } Q^2 +\etaprime Q.
\ee
Now using the equation of motion, one can integrate out $Q$ and get, locally, the $\etaprime$ mass ${\cal L}_Q \to - \frac{1}{2}f_\pi^2 m^2_{{\eta^\prime}} {\etaprime}^2$. Now imposing {\it globally} the $2\pi$ periodicity of $\etaprime$ by taking into account the instanton number $\int Q\in {\cal Z}$, one arrives at the cusp (\ref{cusp}).  It has been argued that when the HLS vector fields are {\it integrated in} as Seiberg-dual to the gluons fields~\cite{komargodski-HLS}, then Chern-Simons theory appears on the $\etaprime=\pi$ sheet (``pancake") encoding the quantum Hall droplet.  And this is found to be encoded in the ``hidden"\footnote{In \cite{HY:PR} without the $\etaprime$ field, ``hidden" is replaced by ``homogeneous."} Wess-Zumino term in the HLS Lagrangian with the chiral field $U$ in which $\etaprime$ field is implemented as 
\be
U=e^{i\pi/f_\pi}\to U=e^{i\eta^\prime}e^{i\pi/f_\pi}.
\ee
This means that the $N_f=1$ baryon is a singular soliton constructed out of $\etaprime$ plus HLS vector mesons in the EFT. 
%
%
%
\section{G$n$EFT and hidden scale symmetry in dense matter}
In baryonic matter under normal conditions, the QH pancake is obviously not ``visible."  If existed at all, it could very well be metastable. So the question is: Starting from a skyrmion matter capturing nuclear matter, under what condition the QH pancake can be arrived at?  
For instance, can one explore whether the QH droplet can be ``seen" at high density by combining HLS with the notion of ``genuine dilaton (GD)" in hidden scale symmetry in nuclear medium?  Scale symmetry was not exploited in \cite{karasik1,karasik2} since it figures as a double-trace contribution which is subleading in $N_c$ in the vacuum. In nuclear medium, however, in particular in the G$n$EFT approach that was used for the EoS of compact-star matter \cite{MR-reviews}, it has a principal role for the properties of neutron stars.

\subsection{``h"Wess-Zumino term in HLS}
To give an idea what may be involved, let me briefly summarize some of the results of \cite{karasik1,karasik2}\footnote{This will be too short in details but gives the key contents that are needed for my arguments.} relevant to my argument and then develop the strategy of how to ``drive" the dense skyrmion system to high density. 
The principal term in the HLS Lagrangian (for flavor $SU(2)$ to which I will be confined) that enters in the discussion is what was referred to as ``h"Wess-Zumio (WZ) term. Here ``h" stands for ``hidden" in  \cite{karasik2} and ``homogeneous" in \cite{HLS-anomaly}. This hWZ term consists of four HLS-gauge-invariant terms that include an external field that correspond to the homogeneous  solution to the  Wess-Zumino anomaly equation. For $N_f=2$ that we are concerned with, the 5D topological term is absent. The hWZ terms conserving parity and charge conjugation but violating intrinsic parity, are written as
\be
\Gamma_{hWZ}=\frac{N_c}{16\pi^2}\int_{M^4} \sum^4_{i=1} c_i {\cal L}_i
\ee
where $c_i$ are arbitrary constants for  the ``homogeneous" WZ terms~\cite{HY:PR} but fixed by Vector Dominance (VD)~\cite{sakurai} for  the ``hidden" WZ terms~\cite{karasik2}. The ${\cal L}_i$s are constructed with Maurer-Cartan 1-forms. The specific forms which can be looked up in \cite{HY:PR,karasik2} are not needed for this paper.  

With the hWZ action analyzed in terms of domain walls, with the HLS vector mesons identified as emergent Chern-Simons fields lodged on the $\etaprime=\pi$ sheet,   it follows that the results can be summarized as follows~\cite{karasik2}. 
\begin{enumerate}
\item  To obtain correct baryon charge by looking at the topological charge and the charge coming from the hWZ term, one must impose 
\be
c_3=c_4=1.\label{c3=c4}
\ee
\item  If one drops the isovector fields $\rho$ and couple $\etaprime$ to the $U(1)$ vector meson $\omega$, then the hWZ action for $N_f=1$ QCD reduces to 
\be
{\cal L}_{hWZ} \to c_3 \frac{N_c}{8\pi^2} d\etaprime \omega d\omega.\label{hWZ}
\ee
Picking the domain wall with $\lim_{z\to -\infty} \etaprime (z)=0$ and  $ \lim_{z\to +\infty} \etaprime (z)=2\pi$, setting $c_3=1$, one gets the $(2+1)$ domain theory
\be
{\cal L}_{DW}=c_3 \frac{N_c}{4\pi} \omega d\omega =  \frac{N_c}{4\pi} \omega d\omega
\ee
which is lodged in the bulk of the sheet.
This relation confirms the $U(1)$ CS theory for the QH pancake giving  the equivalence of $\omega$ to the CS field.
\item Consider $SU(2)$ skyrmions in the presence of $\etaprime$. There is no indication up to date that $\etaprime$ plays any role under normal conditions other than the possible reduction of the $\etaprime$ mass in nuclear medium as discussed above. If QH droplets exist, they must at best be metastable. The situation, more complicated in the presence of the isovector $\rho$ mesons, remains unclear. However the domain wall analysis suggests that the vector mesons lead to a nonabelian CS term as was observed also in the Cheshire Cat analysis~\cite{CCP}. This gives \be
c_1=2/3, \  c_2=-1/3.\label{c1-c2}
\ee
\item The results  (\ref{c3=c4}) and  (\ref{c1-c2}) lead to  VD for all intrinsic parity-violating processes.  These are consistent with nature~\cite{HY:PR}. These results could be taken as (possibly) the first theoretical support for VD~\cite{karasik2}.
\end{enumerate}

Now the important question is: How to see QH droplets and what their role in baryonic matter, if relevant at all, is? My proposal is to ``tweak" $N_f=2$ baryonic matter -- i.e., nuclear matter -- to high density. One way to do this -- that we have exploited profitably to address compact-star matter -- is to combine hidden scale symmetry, intrinsic or emergent, to hidden local symmetry that replaces Chern-Simons field,  leading to QH droplets.
\subsection{Emergent scale symmetry}
To implement hidden scale symmetry (HSS) to hidden local chiral Lagrangian, I take the point of view that it is associated with the ``genuine dilaton (GD)" of Crewther~\cite{GD,CT}. What characterizes this approach is that there is assumed to be an infrared (IR) fixed-point at which scale-chiral symmetry (= scale symmetry plus chiral symmetry) is realized in the Nambu-Goldstone (NG) mode with the pion and  dilaton  NG bosons coexisting with massive matter fields, i.e.,  the HLS mesons and baryons. This is distinct from the conformal window structure being studied in dilatonic Higgs models going beyond the standard model (BSM). A model with a similar IR fixed point realized in the NG mode with  a  ``conformal dilaton" that lies on the border of the conformal window and QCD has been proposed ~\cite{DDZ}.  It is not clear in what way the two dilatons, GD and CD,  are different.  Since the scale symmetry involved in baryonic medium is likely an emergent symmetry, it probably would not matter.  In what follows,  the line of reasoning will follow that of GD. There is no firm confirmation, such as lattice simulation, as to whether the GD (or CD) can be rigorously justified. It will be assumed  in this paper.

Why scale symmetry figures importantly in nuclear dynamics goes as follows. While in matter-free ($n=0$) space, going beyond the single-tracing does not seem warranted for the large $N_c$ limit~\cite{karasik2}.  In the presence of baryonic matter, however, the in-medium VeV proves to be essential. That is because the vacuum changes as density varies. When scale symmetry is implemented, the VeV of the dilaton field $\la\chi\ra$ becomes inevitably density-dependent. This density dependence plays a crucial role in nuclear dynamics as suggested a long time ago~\cite{BR91}. It is a part of what's called ``intrinsic density dependence (IDD)." There are other IDDs coming from matching EFT correlators to QCD ones at, say,  the chiral matching scale $\Lambda_{cs}\sim 4\pi f_\pi\sim 1$ GeV where $f_\pi$ is the pion decay constant~\cite{HY:PR}. It turns out that the density-dependent dilaton condensate $\la\chi\ra^\ast$ (in what follows, unless otherwise stated, $\ast$ stands for density dependence) plays the major role in my arguments.
\subsection{Nuclear effective field theory: G$n$EFT}
Let me briefly define the EFT referred to above as  G$n$EFT that has been successfully exploited in studying the properties of dense star-matter with densities ranging from $\sim n_0$ to $\sim 7 n_0$.  G$n$EFT is a formulation for many-nucleon systems with  scale-chiral symmetric Lagrangian  ${\cal L}_{\psi\chi HLS}$ (here $\psi$ stands for the baryon field, $\chi$ the dilaton field and HLS the hidden vector fields). Up to that density, I assume that QH sheets play no significant role and hence $\etaprime$ can be dropped from the Lagrangian. At what density QH sheets come into the structure of dense baryonic matter cannot be computed in the approach. I will assume it to be beyond the density at which the ``vector manifestation (VM)" of HLS sets in with the gauge coupling $g_\rho\to 0$  and the ``dilaton-limit fixed point (DLFP)" of scale symmetry becomes crucially relevant.  The VM and DLFP are the conditions imposed in G$n$EFT to capture physics from $\sim n_0$ (nuclear) to $\sim 7n_0$ (compact star).

The strategy exploits what was learned in arriving at the compact-star physics in scale-chiral EFT with baryons explicitly included with the Lagrangian ${\cal L}_{\psi\chi HLS}$.  There neither VM nor DLFP is directly probed but their presence was essential. Now the information gained in G$n$EFT with ${\cal L}_{\psi\chi HLS}$ is put into the mesonic Lagrangian ${\cal L}_{\chi HLS}$ with the hWZ terms that capture, as the density exceeds the compact-star,  the $\etaprime$ singularity.  I will argue that  as density increases beyond the neutron-star density, the $SU(2)\etaprime$ system will be  ``driven" to the sheet structure with the QH pancake.

Given the Lagrangian in which HLS and hidden scale symmetry are incorporated, suitably  coupled to baryons ${\cal L}_{\psi\chi HLS}$, it would be ideal to formulate the power series expansion as has been done in N$\chi$EFT and work out N$^p$LO terms (for $p \geq 1$) as is successfully done up to nuclear matter density~\cite{chieft}.  Formulating scale-chiral EFT with {\it only} the pions and the GD dilaton to N$^1$LO exists in the literature~\cite{cata}. The power counting including HLS fields and baryon fields has also been worked out to the same power-counting order~\cite{LMR}. One can see even at the leading order (LO) certain processes, such as $K\to 2\pi$ decay which requires higher order in 3-flavor chiral perturbation theory, are effectively captured at the tree-order in scale-chiral perturbation theory~\cite{CT}. Similarly in nuclear physics, the tree-order dilaton contribution is found to capture high-order $\pi-\pi$ interactions of N$\chi$EFT. It is therefore clear that the role of the dilaton in scale-chiral scheme should have a power that cannot be ignored. Unfortunately, it is at present practically unfeasible to go beyond the LO with ${\cal L}_{\psi\chi HLS}$. There are simply too many  parameters that cannot be controlled.

In the absence of the high-power EFT calculation of the N$\chi$EFT type, the approach in G$n$EFT applied to compact-star physics exploited a different strategy. First the ``genuine dilaton (GD)" field is incorporated in the HLS Lagrangian by means of the ``conformal compensator field (CCF)" $\chi=f_\chi e^{\sigma/f_\chi}$ that transforms linearly both in mass and in scale  (whereas  $\sigma$ identified as the genuine dilaton (GD), transforms  nonlinearly in scale)
\be
{\cal L}_{\rm scale-chiral}={\cal L}^{\rm inv}_{\chi \rm HLS} + V_{D}\label{SCL}
\ee
where ${\cal L}^{\rm inv}_{\chi \rm HLS}$ is HLS- and scale-invariant and $V_{D}$ is the dilaton potential that contains all scale invariance breaking terms such as trace anomaly, quark mass etc. at the leading order (LO) in scale-chiral symmetry in the GD scheme.

The strategy is to take  ${\cal L}_{\rm scale-chiral}$ given to LO, incorporate the nucleon fields explicitly in consistency with HdF symmetries, also to LO, and formulate the many-body system by doing renormalization-group treatments of the baryons on the Fermi surface~\cite{FL}. What is essential for this strategy is that the Lagrangian be defined for the varying ``vacuum" as the nuclear density varies, that is, in the sliding vacua. The sliding figures in the VeVs, such  as $\la\chi\ra^\ast$ and $\la\bar{q}q\ra^\ast$  and in coupling constants such as, particularly, the HLS coupling $g_V^\ast$. This allows EFT Lagrangian, i.e., chiral Lagrangian, to be connected~\cite{MR-Migdal}  to Landau Fermi-liquid theory (for elementary fermions, i.e., both electrons and nucleons) and Landau-Migdal Fermi-liquid theory (for nuclei and nuclear matter)~\cite{migdal}. This procedure elevates the mean-field approximation with the given Lagrangian to Landau Fermi-liquid fixed point theory.  It can be thought as belonging to the class of density functional theory \`a la Hohenberg-Kohn theorem. Systematic corrections can in principle be made as a power series in $1/\bar{N}\sim 1/k_F$ where $\bar{N}=k_F/(\Lambda_{\rm FS}-k_F)$. It has been shown that the Fermi-liquid fixed-point (FLFP) approximation works remarkably well for certain quantities associated with low-energy theorems valid in nuclei~\cite{FR}. As explained in detail in \cite{MR-reviews}, the G$n$EFT in the FLFP approximation works not only at nuclear matter density but also at densities relevant to compact stars (as mentioned,  with so far no known tension with Nature).
\section{G$n$EFT toward the compact-star densities and going beyond}
%
 Logically,  to access the matter with $\FSB$s present, one should start with the baryonic matter described in terms of topological objects with the chiral field $U$ containing $\etaprime$, i.e., skyrmions, and drive the system by increasing density so as to ``expose"  $\FSB$s, another form of topological object. This unfortunately is far from feasible because as mentioned nuclear matter as skyrmions has not been worked out well enough to be useful. To proceed, I will take a two-step process~\cite{Mapping}: First extract from the G$n$EFT  (without $\etaprime$) what are considered to be robust topological properties of skyrmions  relevant to compact-star density,  ``translate" them into  the parameters of the bosonic Lagrangian 
  with $\etaprime$ included denoted as ${\cal L}_{\chi\eta^\prime \rm HLS}$.  The bosonic Lagrangian containing the HLS -- with the hWZ terms-- and the hidden scale symmetry \`a la GD  is then applied to analyze how the solitons given by the Lagrangian  evolve at increasing density above the maximum compact-star density,  $\sim 7 n_0$. In doing this I will ignore possible back-reactions of the $\etaprime$ on the properties of the skyrmions describing the nucleons.\footnote{Taking into account the back-reaction could be an interesting -- and challenging -- problem, as indicated in the analysis of \cite{karasik1}.}

The most important topological information that goes into the G$n$EFT is the topology change that takes place at the density $n_{1/2} \gsim 2n_0$ from skyrmions to half-skyrmions. This is observed in the simulation of skyrmion matter on crystal lattice~\cite{cusp,park-vento}. For sufficiently high $n_{1/2}$, skyrmions on lattice can be justified in the large $N_c$ limit.
The skyrmion-half-skyrmion transition is considered in the approach  to be the putative ``hadron-quark" continuity conjectured in QCD, signaling the change of degrees of freedom. The analysis in compact-star physics puts $n_{1/2}$ at $\sim (2.5 - 3.5) n_0$. 

The principal consequences of this transition that enter in the G$n$EFT are:
\begin{enumerate}
\item\label{first} In the half-skyrmion phase, the chiral condensate is locally non-zero supporting chiral density waves but goes to zero averaged on the lattice, with however non-vanishing pion decay constant. The transition is therefore not a chiral restoration. It remains in the NG (Nambu-Goldstone) mode.
\item In the Skyrme model (with no other fields than pions), a cusp develops from the pion fields in the nuclear symmetry energy ($E_{\rm sym}$)  in the equation of state (EoS) of the baryonic matter. When HLS fields are incorporated the cusp is eliminated, replaced by an inflection, giving a soft-to-hard changeover in the EoS. This gives a simple explanation for the massive compact star mass $M\sim 2 M_\odot$.
\item The role of  the $\rho$ meson field accounts for the nuclear tensor forces as the Higgsed $\rho$ meson mass drops as $\propto f_\pi^\ast$~\cite{HY:PR} up to $\sim n_{1/2}$~\cite{cusp}.\footnote{Let me put this remark as an aside not related directly to the issue addressed. That the skyrmion cusp structure encodes precisely what can be interpreted in conventional nuclear physics for the behavior of the nuclear tensor forces when vector mesons are taken into consideration~\cite{bypetal} is not fully appreciated in the nuclear community where the standard N$\chi$EFT based on Weinberg's seminal approach to nuclear EFT without HdFs has been highly successful at least at normal nuclear matter density. It may appear a bit too unorthodox but the role that the HdFs play in low-energy nuclear processes -- as stressed by Gerry Brown in all of his publications -- as well as here and in compact-star physics in \cite{MR-reviews} attests to the potential power of vector mesons in nuclear dynamics, particularly at the densities higher than the normal.}
\item $f_\pi^\ast$ begins to  stop dropping at  $n\gsim n_{1/2}$ but the gauge coupling $g_\rho$ starts dropping to zero, so the $\rho$ mass goes to zero $m_\rho^\ast\sim g_\rho^\ast\to 0$ with $f_\pi^\ast\neq 0$ as density is increased to $n=n_{VM}\gsim 25 n_0$. This density called ``vector manifestation fixed point (VMFP)"  is dictated by the RG flow for HLS~\cite{HY:PR}. How and where this VMFP sets in  {\it controls} the sound speed in the core of compact stars. For instance, the VMFP cannot be at $\sim (6-7) n_0$, i.e., the star core density~\cite{MR-reviews}.

Up to this point, hidden scale symmetry plays no direct role.  There is no indication either that $\etaprime$ figures.

\item In the GD scheme of hidden scale symmetry, as already mentioned, the IR fixed point supports the NG bosons, the pions, $\pi$, and the dilaton, $\chi$ (assuming the massless $d$ and $u$ quarks) but the matter fields, the baryons $\psi$ and HLS fields $\rho$ and $\omega$ etc., can be massive. This would suggest that the IR-fixed point density $n_{\rm IR}$ could be lower than the VMFP density $n_{\rm VM}$ at which the $\rho$ meson mass is to vanish. I will simply assume this holds. The precise numerical values cannot be given in the theory. 
\item In the Fermi-liquid fixed-point approximation in G$n$EFT, one can take what's called ``dialton limit" that may be reached by high density. For this, one redefines the field $\Sigma=\frac{f_\pi}{f_\chi} U\chi$ with $U=e^{i\pi/f_\pi}\propto \sigma^\prime +i\vec{\tau}\cdot\vec{\pi}^\prime$ and take the limit ${\rm Tr} (\Sigma^\dagger \Sigma)\to 0$. In order to avoid the singularities developing in the limiting process, one is required to put what are referred to as the ``dilaton-limit fixed-point (DLFP)" constraints
\be
f^\ast_\pi\to f^\ast_\chi, \ g^\ast_A\to 1, \ g^\ast_{\rho NN}=g^\ast_\rho g^\ast_{\rm COR} \to 0\label{DLFP}
\ee
where $g^\ast_\rho$ is the HLS gauge coupling in medium and  $g^\ast_{\rm COR}$ stands for nuclear correlation effect in dense matter. {\it It is at this point  the isovector meson $\rho$ decouples from the nucleons as $g^\ast_{\rm cor}\to 0$,  before the gauge coupling $g^\ast_\rho $ drops to zero}. What one winds up with is the Gell-Mann-L\'evy linear sigma model,  gauge-coupled to the $\omega$ meson~\cite{interplay}. { The  DLFP could be close to the IRFP of the GD scheme with $f_\chi^\ast\neq 0$ while $\la\chi\ra^\ast\to 0$. This is consistent with the pseudo-gap property of the chiral condensate in the half-skyrmion phase discussed above.} 
\item\label{last} The DLFP constraints (\ref{DLFP}) have an important consequence that the parity-doubling symmetry {\it emerges} in the nucleon spectra as density goes toward $n_{\rm DLFP}$. This parity-doubling is a consequence of  an extremely intricate interplay between the attraction due to the scalar meson interaction with the nucleons and the repulsion due to the $\omega$ meson. The interplay makes~\cite{interplay} the effective nucleon (quasiparticle) mass approach a density-independent chiral scalar $f_\chi^\ast\propto  m_0$ as density moves toward $n_{\rm DLFP}$. This accounts for the pseudo-conformal sound speed~\cite{PCM} because the trace of the energy-momentum tensor (TEMT) becomes density-independent. The TEMT is given as a function only of $f_\chi^\ast$ in the Landau Fermi-liquid fixed-point approximation -- modulo small corrections verified by $V_{lowk}$RG~\cite{MR-reviews} -- that gives the pseudo-conformal structure from $\gsim  n_{1/2}$ up to $\sim 7 n_0$ in the core of massive stars~\cite{PCM}. 
\item Pushing further up in density beyond the compact-star density,
one arrives at 
\be
\la\chi\ra^\ast\propto f^\ast_\chi \to 0, \la\bar{q}q\ra^\ast\propto f^\ast_\pi\to 0. 
\ee
In this limit, the pseudo-conformal sound speed $v_s^2/c^2\approx 1/3$ found in G$n$EFT~\cite{PCM} will become conformal $v_s^2/c^2=1/3$. This means also that the constant mass $m_0$ will have to vanish.  How this can happen is not precisely known but it seems inevitable at that limit.  The consequence of this observation which will be relevant to what follows is the decay constants $f^\ast_\pi=f^\ast_\chi$  tend to zero,
\be
 f^\ast_\chi=f^\ast_\pi\to 0.\label{Phi}
\ee

\end{enumerate}
\section{Going toward $\FSB$ and back}
I will now exploit what one has learned from G$n$EFT about dense matter near compact-star density (without $\etaprime$ involved) by extrapolating beyond the compact-star densities the various constraints observed in the baryonic interactions with NG bosons and HdFs. In this process,  I may be making somewhat unjustifiable guesses.

The idea is to inject the information gained in the G$n$EFT results  into the  bosonic Lagrangian ${\cal L}_{\chi\eta^\prime \rm HLS}$ (that includes $\etaprime$ and the hWZ term) and treat the Lagrangian for topological structure of baryonic matter at a density beyond the compact-star matter.

The first condition -- and the most crucial one -- that will be exploited is that the $\rho$ field decouples from the baryonic mater before the VM fixed point,  therefore can be dropped from the bosonic Lagrangian.  To the leading order (LO) in G$n$EFT (with $\etaprime$ included), say, G$n\etaprime$EFT, the Lagrangian simplifies to
\be
{\cal L}_{\etaprime\omega\chi}&&= \frac 12 (\del\chi)^2+ V_\chi  + \frac{f_\pi^2}{2} \Phi^2 (\del \etaprime)^2 - \frac 12 \Phi^2 f_\pi^2 m^2_{\etaprime} { \etaprime}^2\nonumber\\
 -&& \frac{1}{4g^2_\omega} (F^2_{\mu\nu}) + \Phi^2 (2f^2_\pi g^2_\omega)\omega^2
+ c_3\frac{N_c}{8\pi^2}d\etaprime \omega d\omega \label{LAG} 
\ee
where $\Phi=\frac{\chi}{f_\chi}$ and $V_\chi$ is the dilaton potential that has a minimum at $\chi=f^\ast_\chi$. The last term in (\ref{LAG}) comes from the hWZ term. Now setting $\chi=f^\ast_\chi+\chi^\prime$, dropping the scalar fluctuations and setting $\Phi^\ast\equiv \frac{f^\ast_\chi}{f_\chi}\to 0$  from (\ref{Phi}), one sees the scalar $\chi$ and vector $\omega$ become massless, leading precisely to the effective Lagrangian (with $c_3=1$ in  ${\cal L}_{hWZ}$, Eq. (\ref{hWZ})) that leads  to  the QH sheet soliton for $\FSB$. Taking into account $U(1)$ gauge invariance which requires edge mode quantization with chiral bosons, it is argued~\cite{karasik1} that the Lagrangian (\ref{LAG}) with the limiting conditions given above imposed reproduces all the results of \cite{komargodski}. 

What this discussion shows is that the complete decoupling of the $\rho$ mesons from the nucleons leads to pure $\FSB$. It does not indicate how the pion and  $\rho$ configurations get ``distorted" as the density is continuously increased to make $f^\ast_\chi$ go to zero from the core density of compact stars -- where $f^\ast_\chi$ is not equal to zero.  The inverse process -- starting from a pure $\FSB$ to a skyrmion -- by means of manipulating the topological configurations could also be analyzed~\cite{karasik1} although how the baryonic matter density does the necessary ``tweaking" is not clear. It would be interesting to map this out in a schematic model.
\section{Further Remarks}

The question that comes up at this point  is how the $\FSB$s get arranged in ultradense matter. One possibility is that the QH sheets are stacked into a layer of ``lasgnes" or  other forms of pasta. In fact analytical (3+1) dimensional dense skyrmions (without $\etaprime$) have been analyzed in the literature in terms of baryonic layers~\cite{canfora}.    Even more interestingly at some high densities, on skyrmion crystal lattice, skyrmions forming a stack of lasagnes, with half-skyrmions living on the sheet, describe the EoS fairly realistically~\cite{park-paeng-vento}. The half-skyrmions on the sheet can be transformed into three 1/3-charged objects, which could ``masquerade" as deconfined quarks in the core of massive compact stars. This appears to resemble what's observed in Nature~\cite{PCM}. This scenario is not totally bizarre however when domain walls are involved. In condensed matter systems, for two vacua, spinons on a domain wall are liberated in a way strikingly similar to domain wall deconfinement of quarks~\cite{wall-deconfinement}. Now in ultradense matter, multi-$\FSB$s could be in the form of stacks of fractional quantized Hall sheets with fractionally charged fermions. It seems highly intriguing that as density increases, the sheets supporting half-skyrmions or perhaps fractional baryons could smoothly go over to the sheets supporting fractional quantum Hall structure. It may be that this movement in hadronic degrees of freedom  going up in density overlaps with the movement going down  in QCD degrees of freedom.  What seems clear is that the cross-over region is full of new physics presently more or less unknown. In fact there are various unforeseen phenomena observed under extreme conditions. For instance there is an indication in lattice QCD at high temperature that there exists a distinct pion state  at $T=220$ MeV, some 1.2 times the chiral (pseudo-)critical temperature $T_{psc}$~\cite{philipsen} and   also a sign of scale invariance in the infrared  at temperatures way above $T_{psc}$~\cite{alexandru}. It seems very likely that dense matter beyond the compact-star density harbors even more surprises.

\end{document}